\documentclass[aip,jcp,reprint]{revtex4-1}
\pdfoutput=1
\usepackage{graphicx}%
\usepackage{dcolumn}%
\usepackage{bm}%
\usepackage{amsmath,amssymb}%
\usepackage{rotating}%
\usepackage{latexsym}%
\usepackage[usenames]{color}%

\DeclareMathOperator{\Tr}{Tr}

\begin{document}

\title{Hydrodynamic Effects on Confined Polymers}

\author{Santtu T. T. Ollila}
\email[e-mail: ]{santtu.ollila@aalto.fi}
\affiliation{COMP Centre of Excellence, Department of Applied Physics, Aalto University School of Science and Technology, P.O. Box 11000, FIN-00076 Aalto, Espoo, Finland}
\affiliation{Department of Applied Mathematics, The University of Western Ontario, London, Ontario, Canada N6A 5B8}
\author{Colin Denniston}
\email[e-mail: ]{cdennist@uwo.ca}
\affiliation{Department of Applied Mathematics, The University of Western Ontario, London, Ontario, Canada N6A 5B8}
\author{Mikko Karttunen}
\email[e-mail: ]{mikko.karttunen@uwaterloo.ca}
\affiliation{Department of Chemistry, University of Waterloo, Waterloo, Ontario, Canada N2L 3G1}
\author{Tapio Ala-Nissila}
\email[e-mail: ]{tapio.ala-nissila@aalto.fi}
\affiliation{COMP Centre of Excellence, Department of Applied Physics, Aalto University School of Science and Technology, P.O. Box 11000, FIN-00076 Aalto, Espoo, Finland}
\affiliation{Department of Physics, Brown University, Providence, Rhode Island 02912-1843 USA }
\date{\today}

%%%%%%%%%%%%%%%%%%ABSTRACT%%%%%%%%%%%%%%%
\begin{abstract}
We consider the statics and dynamics of a flexible polymer confined between parallel plates both in the presence and absence of hydrodynamic interactions. The hydrodynamic interactions are described at the level of the fluctuating, compressible Navier-Stokes equation. We consider two cases: (i) confinement for both the solvent and the polymer, and (ii) confinement for the polymer only (in a 3D solvent), which is experimentally feasible, for instance, by (optical) trapping. We find a continuous transition from 2D to 3D dynamic scaling as a function of decreasing degree of confinement within the de Gennes and the weak-confinement regimes. We demonstrate that, in the presence of hydrodynamics, the polymer's center-of-mass diffusion coefficient in the direction parallel to the walls scales differently as a function of the level of confinement in cases (i) and (ii). We also find that in the commonly used Langevin dynamics description, the polymer swells more parallel to the walls than in the presence of hydrodynamics, and the planar diffusion coefficient shows scaling behavior similar to case (ii) rather than case (i). In addition, we quantify the differences in the static structure factor of the polymer between cases (i) and (ii), and between case (i) and Langevin dynamics.

\end{abstract}
% 05.40.Jc Brownian motion
\maketitle

%%%%%%%%%%%%%%INTRODUCTION%%%%%%%%%%%%%%%%%%
\section{Introduction}
The static and dynamic behavior of polymers in the dilute limit is well understood in the 2D and 3D limits, both of which have been studied theoretically with and without hydrodynamic interactions (HIs)~\cite{doiedwards,deG79,PFVA-N05,Larson05}. The theoretical predictions have been confirmed by experiments~\cite{Larson05,FM93,SPC96,MR99,KF04}. Extensions have been developed to cover, for example, effects of semidiluteness and chain stiffness, and also these theories are well documented and tested~\cite{doiedwards,deG79,N93,GKFBS96,WHR97,LM01}.

Recent developments in promising micro and nanofluidic technologies~\cite{SQ05} have sparked a renewed interest in the statics and dynamics of {\it confined} polymers~\cite{T96,MMO11}. Classic theoretical works on the subject based on scaling theory established the de Gennes and Odijk regimes of confinement~\cite{classicConfined1,*classicConfined2,*classicConfined3,classicConfined4}. The former regime has a polymer confined at least in one dimension into a gap whose size $L$ is smaller than the chain's radius of gyration in bulk, $R_g$, but much greater than its persistence length $l_p$. The de Gennes theory has been corroborated using the Edwards-Singh approach~\cite{CMT97,*CM11}. Odijk developed a theory to describe the relationship between the planar size of the chain $R_\parallel$ and $L$ as $L$ approaches $l_p$.  Motivated by research on DNA, these regimes have been supplemented (with still experimentally unverified relations) to address more intricate regimes of confinement for different types of polymers~\cite{extensionConfined1} and followed by Monte Carlo (MC) simulations without chain-wall electrostatic interactions~\cite{DJMD12,MO12}. Most recently, Dai {\it et al.}~\cite{DJMD12} have done extensive work without HIs on characterizing the transition to the de Gennes regime from weak confinement and from the de Gennes to the Odijk regime. They found $R_\parallel$ to follow a single power law with an exponent of about $1/4$ as a function of $1/L$ from the de Gennes regime up to the onset of the Odijk regime in slitlike confinement. 
Two recent DNA experiments~\cite{BMSD08,UIK10} have measured similar power laws and report exponent values of $\beta=0.23\pm 0.03$ and $0.26\pm 0.04$, respectively.
However, in another recent experimental study, Strychalski {\it al.}~\cite{SGGLS12} found $R_\parallel \sim (1/L)^{\sim 0.16}$ in nanofluidic slitlike confinement close to and in the transition between the de Gennes and Odijk regimes. Moreover, other experimental findings~\cite{SLC08,TLTJCD10} of polymer diffusivity in similar nanofluidic slitlike confinement have contradicted predictions based on blob theory~\cite{classicConfined1,*classicConfined2,*classicConfined3} and suggest the transition from the de Gennes to the Odijk regime to be broad.

The disagreement and controversy between theory and experiments leaves room for numerical simulations that take into account the full gamut of HIs in such quasi-two-dimensional (q2D) systems where the polymer dynamics is strongly restricted in one dimension~\cite{Diamant09}. Particle dynamics is sensitive to hydrodynamic boundary conditions ({\it e.g.}$\!$ porosity, slip or no slip on walls), which can dictate the physical behavior in the system~\cite{Diamant09,OND12}. Thus, the methodology must be chosen carefully for studies of confined suspensions in order not to neglect important features of suspensions dynamics due to limitations in the algorithm~\cite{P04}. Here, we take advantage of a recently developed hybrid Lattice-Boltzmann (LB) model~\cite{ODKA11} that provides a well-defined hydrodynamic radius and boundary conditions for the solute particles~\cite{OND12}.

In the present work, we investigate the transition from 2D to 3D both for static and dynamic scaling of a single polymer confined between two rigid parallel plates. We consider two distinct types of confinement: (i) where both the solvent and the polymer are fully confined (q2D) and (ii) where the solvent is in 3D and only the polymer is constrained by a focusing potential. We quantify the static and dynamic scaling exponents as a function of the degree of confinement. Our method allows us to study the role of no-slip walls on the planar center-of-mass diffusion coefficient of the polymer.

Unlike colloidal suspensions~\cite{CDLR04,*VRCHDL07,*R-SS-SB-CA-L10}, polymer dynamics in the q2D-transitional region with HIs has not been investigated systematically as a function of the degree of confinement. It is not even known conclusively whether static equilibrium quantities are affected by HIs as they are coupled dynamically through fluctuations. Despite the pivotal role of HIs, polymer chains in q2D systems are often studied in equilibrium in the absence of HIs~\cite{DJMD12,MO12}. Full solutions to this complicated hydrodynamic problem are needed in order to provide understanding of fundamental physics in microfluidic experiments~\cite{B-ORS09} and future applications~\cite{SMB11}.

%%%%%%%%%%%%%%THEORY%%%%%%%%%%%%%%%%%%
\section{Theory}
\begin{figure}
\centerline{\includegraphics[width=\columnwidth]{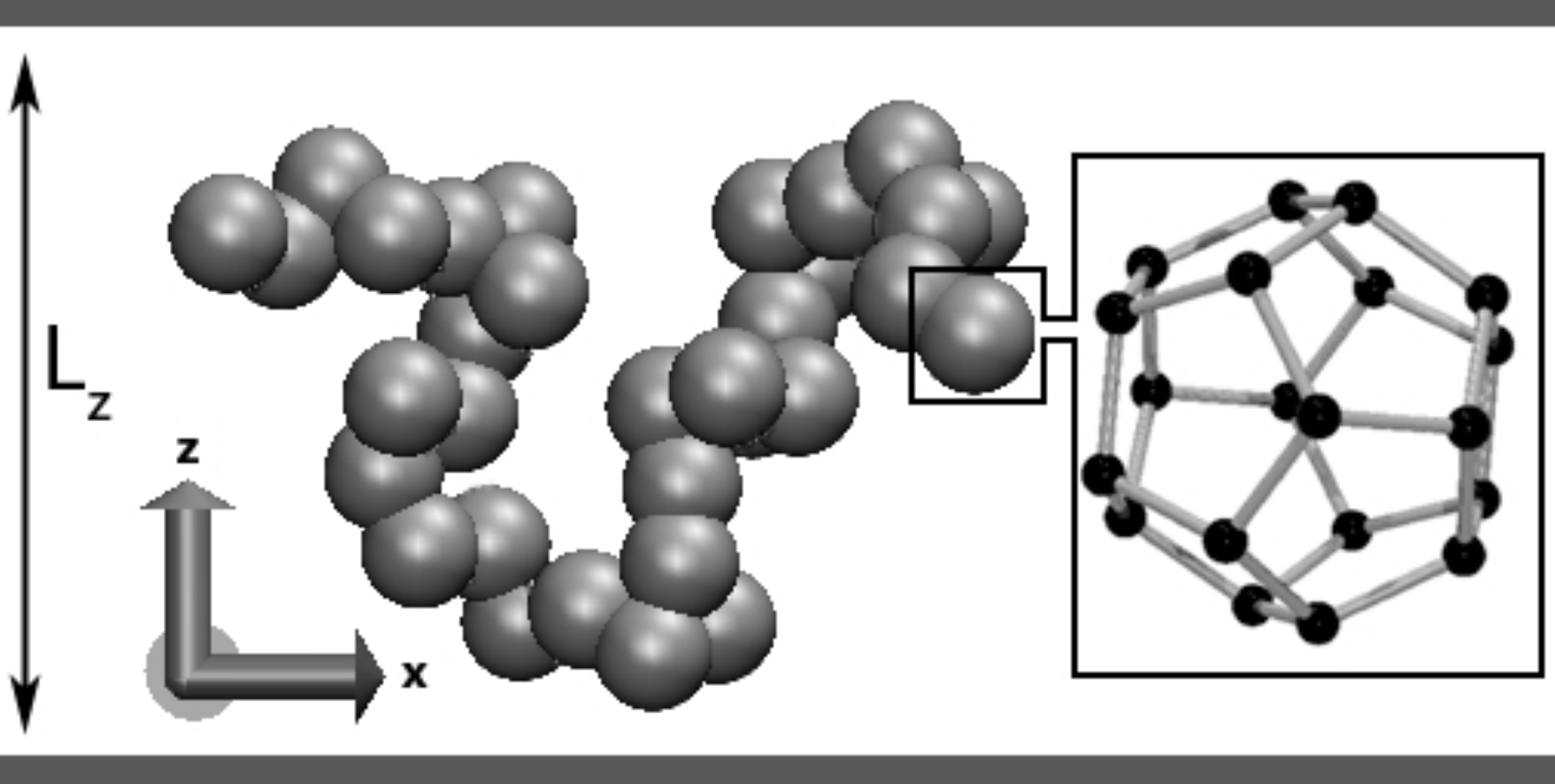}}%
\caption{An illustration of our polymer model and coordinate system in the present work. The frame encloses a schematic of the structure of a $20$-node composite monomer that provides hydrodynamic consistency~\cite{ODKA11}. In this work, the node count is $30$. The picture of the polymer was generated in VMD~\cite{HDS96}.}
\label{fig:schematic}
\end{figure}

In the present work, we shall examine polymers confined between parallel plates a distance $L_z$ apart, as shown schematically in Fig.~\ref{fig:schematic}.  The confinement can result in the polymers adopting an asymmetric configuration which we characterize using the tensor of gyration $Q_{\alpha\beta}$,
\begin{equation}\label{eq:rgt}
Q_{\alpha\beta} = \frac{1}{N}\sum_{n=1}^N (r_{n,\alpha} - r_{\mathrm{cm},\alpha})(r_{n,\beta} - r_{\mathrm{cm},\beta}),
\end{equation}where $N$ is the degree of polymerization, $\alpha,\beta\,\in\,\lbrace x,y,z\rbrace$ and the subscript cm refers to the center of mass of the polymer. The components of $\langle Q \rangle$ will be compared with the isotropic radius of gyration in the absence of confinement, $R_g$, which is related to the trace of Eq.~(\ref{eq:rgt}) and the sum of the tensor's eigenvalues by
\begin{equation}\label{eq:Rgdef}
R^2_g = \langle \Tr Q \rangle = \langle \lambda_1  + \lambda_2 + \lambda_3 \rangle
\end{equation} {\it in free space}. The subscripts of the non-negative eigenvalues indicate their respective magnitude, {\it i.e.} $\lambda_1 \geq \lambda_2 \geq \lambda_3 \geq 0$. The ensemble averages $\langle \lambda_1\rangle/\langle\lambda_2\rangle$ and $\langle \lambda_1\rangle/\langle\lambda_3\rangle$ characterize the anisotropy of the mean shape of the chain. To relate our simulations to experimental imaging of the polymer by a microscope, we define the size of the polymer parallel, $R_\parallel$, and normal, $R_\perp$, to confining plates through
\begin{equation}\label{eq:Rcomponents}
R^2_{\parallel} \equiv \Bigl\langle \sum_{i=1}^3  \lambda_i(1-\vert\hat{{\bf e}}_i \cdot \hat{{\bf z}}\vert^2)\Bigr\rangle;\quad R^2_{\perp} \equiv \Bigl\langle\sum_{i=1}^3 \lambda_i \vert\hat{{\bf e}}_i \cdot \hat{{\bf z}}\vert^2 \Bigr\rangle,
\end{equation}where $\hat{{\bf e}}_i$ are the three unit-normalized eigenvectors corresponding to $\lambda_i$ and $\hat{{\bf z}}$ is a unit vector perpendicular to the confining planes. We shall examine the polymer between parallel plates separated by a distance $L_z$, which we use together with Eq.~(\ref{eq:Rgdef}) to define the degree of confinement, $C$, as
\begin{equation}
\label{eq:C}
C = \frac{R_g}{L_z},
\end{equation}where $R_g$ is the free-space radius of gyration. As the Odijk regime with HIs is not feasible for our numerical method, we concentrate on the de Gennes scaling regime, $1 < C \ll R_g/l_p $, where $l_p$ is the persistence length of the chain. For this range of $C$, we recapitulate how the polymer size parallel to the plates, $R_\parallel$, should scale as a function of $C$. De Gennes proposed~\cite{deG79} that between the plates, the self-avoiding polymer forms a set of sequentially connected blobs of linear size $L_z$ each of which contains $g$ monomers. The part of the contour contained within a blob behaves like a self-avoiding walk (SAW), {\it i.e.} $L_z \sim \sigma g^\nu$, where $\nu=0.588$ is the asymptotic value of the 3D Flory exponent for an unconstrained SAW at infinite dilution~\cite{LMS95}. Finally, the blobs behave like a 2D SAW for which reason the planar size of the confined chain scales as
\begin{align}\label{eq:RparavsC}
R_\parallel & \sim L_z (N/g)^{\nu_{2D}} \sim L_z N^{\nu_{2D}} (\sigma/L_z)^{{\nu_{2D}/\nu}}\nonumber\\
& \sim L_z N^{\nu_{2D}} \left(\frac{\sigma N^\nu}{L_z N^\nu}\right)^{{\nu_{2D}/\nu}}\\
\Rightarrow \frac{R_\parallel}{R_g} & \sim C^{\nu_{2D}/\nu-1} \equiv C^\beta,\nonumber
\end{align}where we have used the Flory radius of gyration~\cite{Flory69} $R_g \sim \sigma N^\nu$. The long-chain limit for the exponent $\beta = \nu_\mathrm{2D}/\nu-1 \approx 0.276$, where $\nu_{2D}=3/4$ is the asymptotic value of the Flory exponent in 2D~\cite{deG79}. It is important to remember that in actual numerical simulations~\cite{ODKA11} with finite $N$ the value of $\nu$ ($\nu_\mathrm{2D}$) is larger than $0.588$ ($0.75$) due to finite-size effects.

The equilibrium size of a polymer is related to the static structure factor~\cite{doiedwards}, defined as
\begin{equation}\label{eq:Sk}
S({\bf k}) = \frac{1}{N}\sum_{m,n=1}^N \Bigl\langle \exp [\,i\, {\bf k} \cdot ( {\bf r}_m - {\bf r}_n ) ]\Bigr\rangle.
\end{equation}The structure factor of an unconfined self-avoiding chain at infinite dilution scales as $S(k) \sim k^{-1/\nu}$ in the scaling region $ k R_g \gg 1$ ($k=\vert {\bf k} \vert$). These scaling results are subject to finite-size corrections from which asymptotic behavior can be extracted.~\cite{ODKA11}

The dynamic version of Eq.~(\ref{eq:Sk}) defines the intermediate scattering function (or, the dynamic structure factor) $S({\bf k},t)$,
\begin{equation}\label{eq:Skt}
S({\bf k},t) = \frac{1}{N} \sum_{m,n=1}^N \Bigl\langle \exp [\, i\, {\bf k} \cdot (\tilde{{\bf r}}_m(t+s) - \tilde{{\bf r}}_n(s)) ]\Bigr\rangle_s,
\end{equation}where $\tilde{{\bf r}}(t) = {\bf r}(t) - {\bf r}_\mathrm{cm}(t)$. The subtraction of the center-of-mass (CM) motion in Eq.~(\ref{eq:Skt}) is important in order to extract scaling for the intramolecular dynamics only.~\cite{MRWG05} Also, most close-range contributions to the sum in the analogue to Eq.~(\ref{eq:Skt}) in the laboratory frame are between consecutive monomers along the backbone, which results in scaling corresponding to semi-flexible chains in the laboratory frame of reference.~\cite{WHR97,MRWG05,ODKA11} The scaling prediction for Eq.~(\ref{eq:Skt}) is~\cite{doiedwards}
\begin{equation}\label{eq:Sktscaling}
S(k,t) = S(k,0) F(k^z t),
\end{equation}where $F(k^z t)$ is a scaling function and $z=2+\nu_D/\nu$ is the dynamic scaling exponent. Equation~(\ref{eq:Skt}) is expected to hold at infinite dilution for a long polymer at intermediate lengths~\cite{doiedwards} $k\,\in\,(2\pi/R_g,2\pi/\sigma)$, where $\sigma$ is the size of the monomer. The quantity $\nu_D$ is the scaling exponent associated with how the center of mass diffusion coefficient of the polymer, $D$, scales as a function of $N$: $D \sim N^{-\nu_D}$. In 2D, it is known that $\nu_D=0$ in the presence of hydrodynamic interactions as $D \sim \log N$ for which reason $z=2$.~\cite{PFVA-N05,FPVA-N03} In 3D, $\nu$ and $\nu_D$ are equal according to the Zimm model resulting in $z=3$. Self-avoiding chains {\it in the absence of hydrodynamic interactions} have $\nu_D=1$ for dilute 3D systems and for all polymer concentrations in 2D leading to $z \approx 3.7$ (3D) and $z=10/3$ (2D).

We analyze the static and dynamic structure factors separately parallel to confining planes, $S_\parallel$, and perpendicular to them, $S_\perp$, which we define as
\begin{subequations}
\label{eq:Sktconf}
\begin{eqnarray}
S_\parallel(k,t) & = & S({\bf k}_\parallel,t) = S(k({\bf e}_x+{\bf e}_y)/\sqrt{2},t);\label{eq:Sktconf:a}\\
S_\perp(k,t) & = & S({\bf k}_\perp,t)= S(k{\bf e}_z,t)\label{eq:Sktconf:b}.
\end{eqnarray}
\end{subequations} We expect $S_\parallel$ to transition from the known 3D to 2D result with increasing level of confinement in the $z$ direction. The case of $S_\perp$ is not as clear at high degree of confinement and we do not expect it to exhibit scaling.

%%%%%%%%%%%%%%MODEL%%%%%%%%%%%%%%%%%%
%%%%%%%%%%%%%%MODEL%%%%%%%%%%%%%%%%%%
%%%%%%%%%%%%%%MODEL%%%%%%%%%%%%%%%%%%
\section{Model}
\subsection{Geometry}
The system is bounded by two parallel plates spanning a planar region $[0,L_x] \times [0,L_y]$ located at coordinates $z_0=\Delta x$ and $N_z\Delta x$, where $N_z$ is a number of fluid lattice sites chosen to give a desired degree of confinement. Figure~\ref{fig:schematic} contains a schematic side view of the system. Periodic boundary conditions (PBCs) are applied in the $x$ and $y$ directions with $L_x=L_y$ and they are chosen to be about $10 R_g$, which we find sufficient for eliminating finite-size effects in Section~\ref{sec:monomdist}. The polymer is confined between the plates by a shifted and truncated 12-6 Lennard-Jones (LJ) potential normal to the planes. At the bottom wall this is
\begin{equation}\label{eq:Umw}
U_\mathrm{mw}=4\epsilon\left(\Bigl(\frac{\sigma_\mathrm{mw}}{z-z_0}\Bigr)^{12}-\Bigl(\frac{\sigma_\mathrm{mw}}{z-z_0}\Bigr)^6 + \frac{1}{4}\right),
\end{equation}where the energy scale is set by $\epsilon = k_\mathrm{B} (300\,\mathrm{K})$, $\sigma_\mathrm{mw}=1.87\,\sigma$, $k_\mathrm{B}$ is the Boltzmann constant and $\sigma=1.5\,\mathrm{nm}$. The interaction is cut off at $2^{1/6}\sigma_\mathrm{mw}\approx 2.1\,\sigma$. An analogous potential is used at the top.  We define the plate separation in Eq.~(\ref{eq:C}) as
\begin{equation}
L_z = (N_z\Delta x - \sigma_\mathrm{mw})-(\Delta x+\sigma_\mathrm{mw}).
\end{equation}

\subsection{Polymer}
We use the finitely extensible nonlinear elastic (FENE) chain~\cite{GK86} to model a linear polymer as $N$ monomers whose centers of mass (CMs) are connected by $N-1$ bonds. To capture the excluded-volume effect, we apply the shifted and truncated 12-6 LJ potential between the CMs of all pairs of monomers. These two features can be written for a pair of particles $i,j$ as
\begin{eqnarray}
&& U({\bf r}_i,{\bf r}_j) = -\frac{1}{2} k R^2_0 \log \left(1- \frac{r^2_{ij}}{R^2_0}\right)\nonumber\\
&&+ 4\epsilon\left(\Bigl(\frac{\sigma}{r_{ij}}\Bigr)^{12}-\Bigl(\frac{\sigma}{r_{ij}}\Bigr)^6+ \frac{1}{4}\right)\Theta(2^{1/6}-r_{ij}/\sigma)\label{eq:fenelj},
\end{eqnarray}
where $r_{ij}=\vert {\bf r}_i - {\bf r}_j\vert$, the first term is the FENE bond present only for consecutive beads of the chain, $\Theta(x)$ is the Heaviside step function and the second term is the repulsive LJ potential. The parameters of the model are the spring constant $k=32.6\epsilon \sigma^{-2}$ and the maximum extension of a bond $R_0=1.5\sigma$. Each monomer $i$ follows Newton's equation of motion
\begin{equation}
m\ddot{{\bf r}}_i = - \displaystyle\sum_{j\neq i} \nabla_i U({\bf r}_i,{\bf r}_{j}) + {\bf F}_i({\bf r}_i,{\bf v}_i,{\bf u}({\bf r}_i,t)),
\end{equation}where the mass of the unsolvated particle is $m=24 \times 10^{-24}\,\mathrm{g}$. The second term on the right-hand side is a local force to describe frictional fluid-particle interaction,
\begin{eqnarray}
{\bf F}_i &=& \int_{B_i(t)} -{\bf f}_i\,\mathrm{d}^3 x\nonumber\\
&=& \!-\! \int_{B_i(t)}\!\!\gamma\, n_i({\bf r})\Bigr(\!{\bf v}_i\!+\!{\bf w}_i \! \times \! ({\bf r} - {\bf r}_i) \!-\! {\bf u}({\bf r})\!\Bigl)\mathrm{d}^3 x,\label{eq:ourFf}
\end{eqnarray}where ${\bf w}_i$ is the monomer's angular velocity and ${\bf v}_i$ its center-of-mass (CM) velocity. The local fluid velocity ${\bf u}({\bf r})$ that enters Eq.~(\ref{eq:ourFf}) provides correct thermalization for the chain without having to add external noise to the particles when $\gamma$ is chosen appropriately~\cite{ODKA11,MacKayD12}. The composite monomer of volume $B_i$ consists of $30$ nodes arranged on a roughly spherical shell as shown in the inset of Fig.~\ref{fig:schematic} at a distance of $R=0.7\,\mathrm{nm}$ about the CM coordinate ${\bf r}_i$. By modeling a monomer in this fashion, correct thermalization and consistency between different measures of the monomer's hydrodynamic radius is attained by choosing the coupling constant $\gamma$ as detailed in Ref.~\onlinecite{ODKA11}.

\subsection{Solvent}
\subsubsection{Lattice-Boltzmann fluid}
Our solvent model reproduces equations for the mass and momentum conservation in a fluid at the Navier-Stokes level that read
\begin{equation}
\partial_t \rho + \partial_{\alpha}(\rho u_{\alpha}) = 0
\label{eq:continuity}
\end{equation}and
\begin{align}
&\partial_t(\rho u_\alpha) + \partial_\beta(\rho u_\alpha u_\beta) =  -\partial_\alpha P_{\alpha\beta} + f_\alpha\label{e:NavierStokes}\\
&\,+\partial_\beta\left(\eta\Bigl(\partial_\alpha u_\beta+\partial_\beta u_\alpha - \frac{2}{3}\partial_\gamma u_\gamma\delta_{\alpha\beta}\Bigr) +  \zeta \partial_\gamma u_\gamma\delta_{\alpha\beta}\right),\nonumber
\end{align}
where $\rho$ and $u_\alpha$ are the fluid density and components of velocity, $\eta$ and $\zeta$ are the shear and bulk viscosities and $P_{\alpha\beta}$ is the fluid pressure. In this work, the pressure tensor is diagonal with linear dependence on density, {\it i.e.} $P_{\alpha\beta} = \rho v^2_s \delta_{\alpha\beta}+s_{\alpha\beta}$, where $v_s$ is the speed of sound and $s_{\alpha\beta}$ is related to thermal fluctuations~\cite{ODKA11}. This can be viewed as an ideal gas equation of state or the first term in a Taylor expansion of the pressure about fixed density in which case $v^2_s$ is the isentropic compressibility~\cite{Kell70}.  The force density exerted by the polymer appears through $f_\alpha$. Our standard LB fluid algorithm reproduces Eqs.~(\ref{eq:continuity}) and (\ref{e:NavierStokes}) in the form typical to most LB algorithms.~\cite{S01,*CD98} The shear viscosity in the model is $\eta = \rho (\tau-\Delta t/2) v_c^2/3$, where $v_c=\Delta x/\Delta t$ is a lattice velocity, and $\zeta = \eta(5/3-3v_s^2/v_c^2)$.~\cite{SOY95} In this paper, $\tau=1.13\Delta t$ and $\rho = m/\Delta x^3$. The speed of sound $v_s$ is chosen to be $v_s=v_c/\sqrt{3}$ ($v_s<v_c$ is required for stability in LB algorithms). Thermal fluctuations in the fluid stress tensor $s_{\alpha\beta}$, and corresponding noise in higher moments, maintain a constant temperature of $T=300\,\mathrm{K}$.~\cite{ODKA11} Within the simulations themselves, computation is done in units of the lattice discretization $\Delta x=1.0\,\mathrm{nm}$ and $\Delta t=0.15\,\mathrm{ps}$. The fluid boundaries are located half a lattice constant outside the MD walls, at $z=\Delta x/2$ and $z=(N_z+1/2)\Delta x$, and they are implemented using the mid-grid bounceback rule~\cite{S01}. $N_z$ is the number of lattice sites in the $z$ direction.

\subsubsection{Langevin dynamics}
We also ran Langevin dynamics (LD) simulations of a chain of point particles in which the LB coupling was completely absent. We set the point monomer mass to $2m$. The fluid-particle interaction of Eq.~(\ref{eq:ourFf}) for the LD runs was simply replaced by the common combination of a frictional force and a Gaussian random force uncorrelated in space and time,
\begin{subequations}
\label{eq:LangF}
\begin{eqnarray}
F_\alpha & = & - \xi v_\alpha + F^\mathrm{R}_\alpha;\label{eq:LangF:a}\\
\langle F^\mathrm{R}_\alpha \rangle & = & 0;\label{eq:LangF:b}\\
\langle F^\mathrm{R}_\alpha({\bf r}_i,t) F^\mathrm{R}_\beta({\bf r}_j,t^\prime) \rangle & = & 2 k_\mathrm{B} T \xi\times\nonumber\\
&&\delta({\bf r}_i - {\bf r}_j)\delta(t-t^\prime)\delta_{\alpha\beta}\label{eq:LangF:c}.
\end{eqnarray}
\end{subequations}The friction in the Langevin equation was set to $\xi = 6 \pi \eta R$, where $R$ is the hydrodynamic radius of the monomers in the LB simulations. The time step in Langevin simulation was set to $\Delta t = 35\,\mathrm{fs} \approx 0.007\tau_\mathrm{LJ}$, where $\tau_\mathrm{LJ} \equiv \sigma\sqrt{2m/\epsilon}$.

%%%%%%%%%%%%%%RESULTS%%%%%%%%%%%%%%%%%%
\section{Results}

\subsection{Statics}
The size and shape of a long polymer in a slit has been studied recently using Monte Carlo simulations~\cite{DJMD12,MO12}. These studies had extensive sampling even for chains of several hundred monomers. However, hydrodynamic interactions were not included. We have simulated chains of $N=32$ to $96$ monomers with (LB) and without (LD) hydrodynamics. Our study spans from weak confinement, $C \approx 0.2$, to the de Gennes regime, $1 < C \ll R_g/l_p $, where $l_p$ is the persistence length of the chain.

\subsubsection{Monomer distribution}
\label{sec:monomdist}
We start by examining how monomers are distributed in the slit by looking at the in-plane distribution of monomers $P_\parallel (r)$ about the chain's center of mass, where $r=\langle \sum_n\vert {\bf r}_{n,\parallel} - {\bf r}_{\parallel,\mathrm{cm}}\vert/N\rangle$, and ${\bf r}_{n,\parallel}=(r_{n,x},r_{n,y})$ is the in-plane location of the $n$th monomer. The no-slip walls destroy momentum conservation in the $z$ direction and reduce hydrodynamic finite-size effects~\cite{LM76} from $1/L$ to $1/L^2$, which is a consequence of the form of pairwise HIs (mobility tensor) in q2D systems~\cite{CDLR04,Diamant09}.

\begin{figure}
\centerline{\includegraphics[width=1.0\columnwidth]{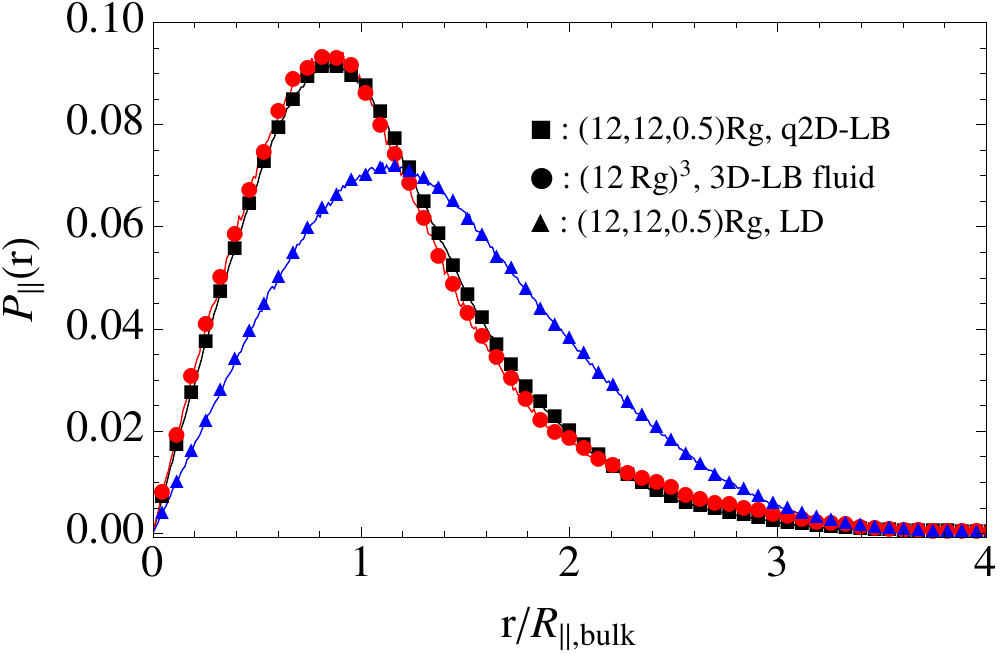}}%
\caption{Probability distribution $P_\parallel(r)$ of the planar monomer-CM distance $r=\langle \sum_n\vert {\bf r}_{n,\parallel} - {\bf r}_{\parallel,\mathrm{cm}}\vert/N\rangle$ for $(N,C)=(96,2.0)$. The abscissa in the figure is scaled by the planar free-space radius of gyration of the chain. The legend refers to the LB and LD box sizes, see text for details.}
\label{fig:monomerdist}
\end{figure}

We examine three cases.  In all of them $(N,C)=(96,2.0)$ and the potential of Eq.~(\ref{eq:Umw}) confines the polymer between $z_0=\Delta x$ and $N_z \Delta x$. In the first case, the LB fluid box size is $(L_x,L_y,L_z) \approx (12,12,0.5)R_g$ (squares in Fig.~\ref{fig:monomerdist}).  No-slip walls for the fluid are in place at the top and bottom of the system. As polymers can be trapped experimentally in slit-like confinement either by walls acting on both the polymer and solvent or by a potential acting solely on the polymer ({\it e.g.} via an optical trap), we also study the second type of confinement. Therefore, in the second case, the LB fluid box is of size $(12 R_g)^3$ (circles) and has periodic boundary conditions (PBCs) in all directions. This changes the form of HIs from q2D (squares) to 3D (circles). We also include Langevin dynamics (LD) in $(L_x,L_y,L_z) \approx (12,12,0.5)R_g$ as the third case (triangles). The resulting $P_\parallel (r)$ for each case is shown in Fig.~\ref{fig:monomerdist}.

The planar radius of gyration, $R_{\parallel}$, is related to $P_\parallel (r)$ via
\begin{equation}
R^2_{\parallel}=\int_0^\infty r^2 P_\parallel (r)\,\mathrm{d}r.
\end{equation} 
This formula gives $R_{\parallel}=1.18 R_{\parallel,bulk}$ (squares), $1.19 R_{\parallel,bulk}$ (circles) and $1.41 R_{\parallel,bulk}$ (triangles) for the three cases mentioned above. These values, and the plot of $P_\parallel (r)$ shown in Fig.~\ref{fig:monomerdist}, indicate the monomer distribution for the 3D-LB fluid to be much like that for the q2D-LB fluid (squares vs. circles). The LD chain extends $23\%$ more ($R_g(\mathrm{LD})/R_g(\mathrm{LB}) \approx 1.23$) at this level of confinement. Albeit weak, we note that the joint effect of no-slip walls (see text and Fig.~\ref{fig:zmonomerdist} below) and the form of HIs in q2D is observable for a polymer at a level of confinement $C \approx 2$ as a slightly fatter tail (squares vs. circles in Fig.~\ref{fig:monomerdist}). Even though the data is smooth, the slightly fatter tail in q2D than in 3D cannot be attributed directly to an effective pairwise antidrag in q2D between monomers~\cite{CDLR04,Diamant09} since the slight difference occurs around $r\approx 1.5 R_{\parallel,\mathrm{bulk}} \approx 2.3 L_z$. This is not in the limit $r \gg L_z$ where q2D interactions should hold for blobs in de Gennes' theory. Going from an unconfined fluid (3D-LB, circles) to q2D (squares) changes the form of long-range interparticle interaction from $1/r$ to $1/r^2$, which also reduces finite-size effects due to self-interactions across periodic boundaries. We found finite-size effects to be negligible by simulating $(N,C)=(32,1.6)$ for $L_x=L_y=10 R_g$ to $20 R_g$ and saw no discernible changes in $P_\parallel(r)$ between them.
 
We will also later compare our results to LD simulations.  A precise quantitative comparison between LD and LB simulations can be problematic even if the MD parameters are equal as the presence of HIs and no-slip boundary condition modifies the level of confinement in the $z$ direction. This can be seen in the probability distribution $P_\perp (z)$ (normalized as $\int_{z_0}^{N_z \Delta x} P_\perp(z)\,\mathrm{d}z = 1$) of monomer position between the plates shown in Fig.~\ref{fig:zmonomerdist}. Monomers on the Langevin chain (triangles) are able to get closer to the MD wall than those in a q2D-LB (circles) or a 3D-LB fluid (squares). The no-slip walls in q2D provide additional repulsion compared to the 3D-LB fluid (circles vs. squares).

\begin{figure}
\centerline{\includegraphics[width=1.0\columnwidth]{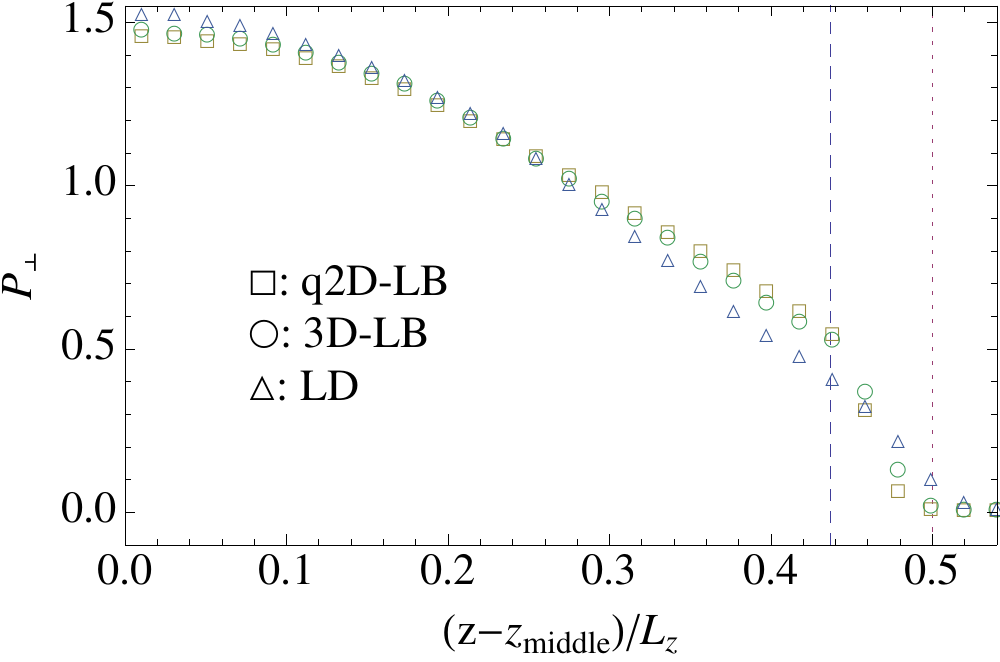}}%
\caption{Half of the symmetric probability distribution $P_\perp (z)$ of monomer positions between confining plates as a function of scaled distance from the center of the channel for a $96$-bead chain for q2D-LB (squares), 3D-LB (circles) and Langevin dynamics (triangles) at $C=2.0$.  The dashed line is located at the cutoff ($2^{1/6}\sigma_\mathrm{mw}$ from the wall) of monomer-wall interaction and the dotted line is at $\sigma_\mathrm{mw}$ from the wall, see Eq.~(\ref{eq:Umw}). The thinner tail of the LD distribution as compared to q2D-LB tells how for a given $C$, the two types of dynamics yield effectively different degrees of confinement.}
\label{fig:zmonomerdist}
\end{figure}

\subsubsection{Radius of gyration}
\begin{figure}
\centerline{\includegraphics[width=0.99\columnwidth]{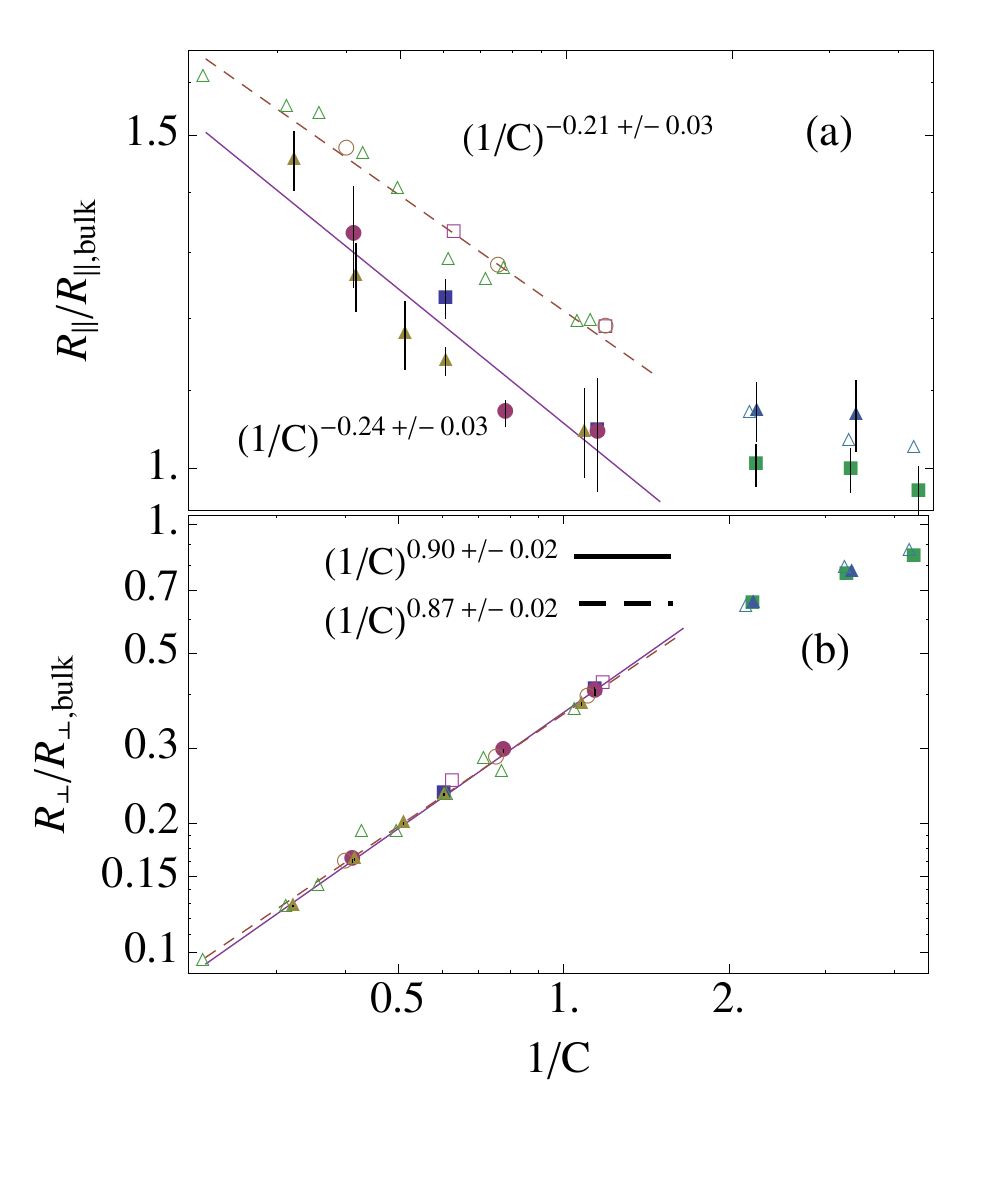}}
\caption{Normalized (a) planar and (b) perpendicular size of the polymer in log-log scale as a function of the inverse of the degree of confinement for LD simulations without HIs (hollow symbols) and q2D-LB simulations with HIs (solid symbols). The MD confining plates according to Eq.~(\ref{eq:Umw}) are identical in all cases. Data points are for $N=32$ (large squares), $64$ (large circles) and $96$ (up triangles) so $1/C=\mathrm{const.}$ may correspond to different $R_g$-$L_z$ combinations. The symbols corresponding to $1/C > 2$ are excluded from the fits whose results are shown as straight lines for q2D-LB data and dashed lines for LD data.}
\label{fig:Rg}
\end{figure}

Figure~\ref{fig:Rg} shows the normalized planar, $R_{\parallel}$, and perpendicular, $R_{\perp}$, radius of gyration as a function of $1/C$ for both q2D-LB (solid symbols) and LD (hollow symbols) simulations. We find significant deviations in $R_{\perp}$ and $R_\parallel$ from bulk behavior for $1/C\leq 2$. Linear fits to the logarithm of the data yield scaling exponents $\beta = 0.24\pm0.03$ for q2D-LB and $0.21\pm0.03$ for LD. The scaling of the perpendicular component $R_\perp$ in Fig.~\ref{fig:Rg} is close to linear indicating that for $1/C=L_z/R_g < 2$ the chain takes up all the available space in the $z$ direction independent of the type of dynamics. The LD chain has a larger relative extension in the parallel component $R_\parallel$ compared to the LB chain. We attribute this to the absence of solvent-mediated correlations and the missing resistance to monomer movement perpendicular to the backbone of the chain. Both data sets (q2D-LB and LD) contain points for $N=32$, $64$ and $96$. In the perpendicular direction there is very little difference between q2D-LB and LD results as the chain takes up all available space in that dimension. However, the approach to unity is much slower than in the parallel direction.

Quantitative comparisons to the theoretical prediction of Eq.~(\ref{eq:RparavsC}) have been performed in recent years both experimentally~\cite{BMSD08,UIK10,TLTJCD10,SGGLS12} and by computer simulations  with~\cite{CGdePRGD04,TLTJCD10} and without HIs~\cite{DJMD12,MO12}. We note that some publications define $C$ in Eq.~(\ref{eq:RparavsC}) as $R_{\parallel,\mathrm{bulk}}/L_z$ instead of according to Eq.~(\ref{eq:C}). This does not affect the scaling exponent as the relation $R_{\parallel,\mathrm{bulk}} = \sqrt{2/3} R_g$ holds in free space due to isotropy.

We comment on the results of computer simulations first. The value of $\beta$ depends on the 3D Flory exponent, which itself depends on $N$ due to finite-size effects for the range of $N$ considered here. We have observed~\cite{ODKA11} values from $\nu(N=32)=0.628\pm0.004$ to $\nu(N=96)=0.595\pm0.003$ in bulk with HIs, from which we could extract the value $\nu=0.586\pm0.005$ for the asymptotic $N\rightarrow \infty$ limit. Dai {\it et al.} used $N=101$ to $401$-bead chains at $C \in (0.1,37)$ in their Monte Carlo study~\cite{DJMD12} of slitlike confinement. In order to define bounds for the de Gennes regime, they chose $L_z \in [2 l_p,R_{\parallel,\mathrm{bulk}}]$, which kept deviations between data points and the line of best fit below $6\%$ for $C\in (0.6,3.1)$. These criteria yielded power law exponents of $\beta(N=101)=0.236 \pm 0.023$ to $\beta(N=401)=0.255 \pm 0.003$, which translate to $\nu(N=101) = 0.607\pm0.011$ and $\nu(N=401)=0.598\pm0.001$ with $\nu_{2D} = 3/4$ using Eq.~(\ref{eq:RparavsC}). We note that also the value of $\nu_{2D}$ is larger than $3/4$ for finite $N$, which, if used in calculating $\nu(N)$, would only increase the value of $\nu(N)$ based on a given $\beta$. Dai and coworkers~\cite{DJMD12} went further by studying the shift from the de Gennes to the Odijk regime, but this was carried out using the effective width of the chain as the scaling parameter which would correspond to $C > R_g/(2 l_p)$ in our model. However, the width parameter is not included in the our model for which reason we cannot access the Odijk regime in the present work. Micheletti and Orlandini~\cite{MO12} found $\beta=0.27 \pm 0.05$ for $C\,\in\,[2,5]$ and $N\,\in\,[360,480]$ in their MC work on semi-flexible chains. Our LD simulations for $32$, $64$ and $96$-bead chains yield $\beta = 0.21\pm0.03$ (hollow symbols in Fig.~\ref{fig:Rg}). Also, our result $\beta = 0.24\pm0.03$ based on q2D-LB data agrees well with the result $\beta\approx0.24$ from an early Brownian Dynamics study~\cite{CGdePRGD04}. Guided by these numbers, it is fair to state that computer simulations of coarse-grained flexible and semi-flexible polymers both with and without HIs, considering the finite value of $N$, support de Gennes' theory of Eq.~(\ref{eq:RparavsC}) if $1\lesssim C \ll R_g/l_p$. However, due to the limited range of values of $C$ that span less than a decade in the present and previous simulations, it is impossible to conclusively verify the existence of the de Gennes scaling regime.

Two recent DNA experiments~\cite{BMSD08,UIK10} report values of $\beta=0.23\pm0.03$ ($2.25\leq C \leq 7.8$) and $0.26\pm0.04$ ($0.4 \leq C \leq 1.7$), respectively. However, other recent experiments~\cite{TLTJCD10,SGGLS12} probe levels of confinement up to $C\approx 20$ and $C\approx 132$ and suggest lower values of $\beta=0.17$ (based on figure 4a in Ref.~\onlinecite{TLTJCD10}) and $\beta=0.16\pm0.01$, respectively. We note that the onset of the theoretical Odijk regime ($L_z \approx l_p$, where $l_p$ is the chain's persistence length) corresponds to $12 < C < 20$ for double-stranded $\lambda$-DNA~\cite{BMSD08}.  Such high values of confinement are beyond the scope of the type of model we consider here which loses its physical relevance in the Odijk regime.

\subsubsection{Static structure factor}
\begin{figure}
\centerline{\includegraphics[width=1.0\columnwidth]{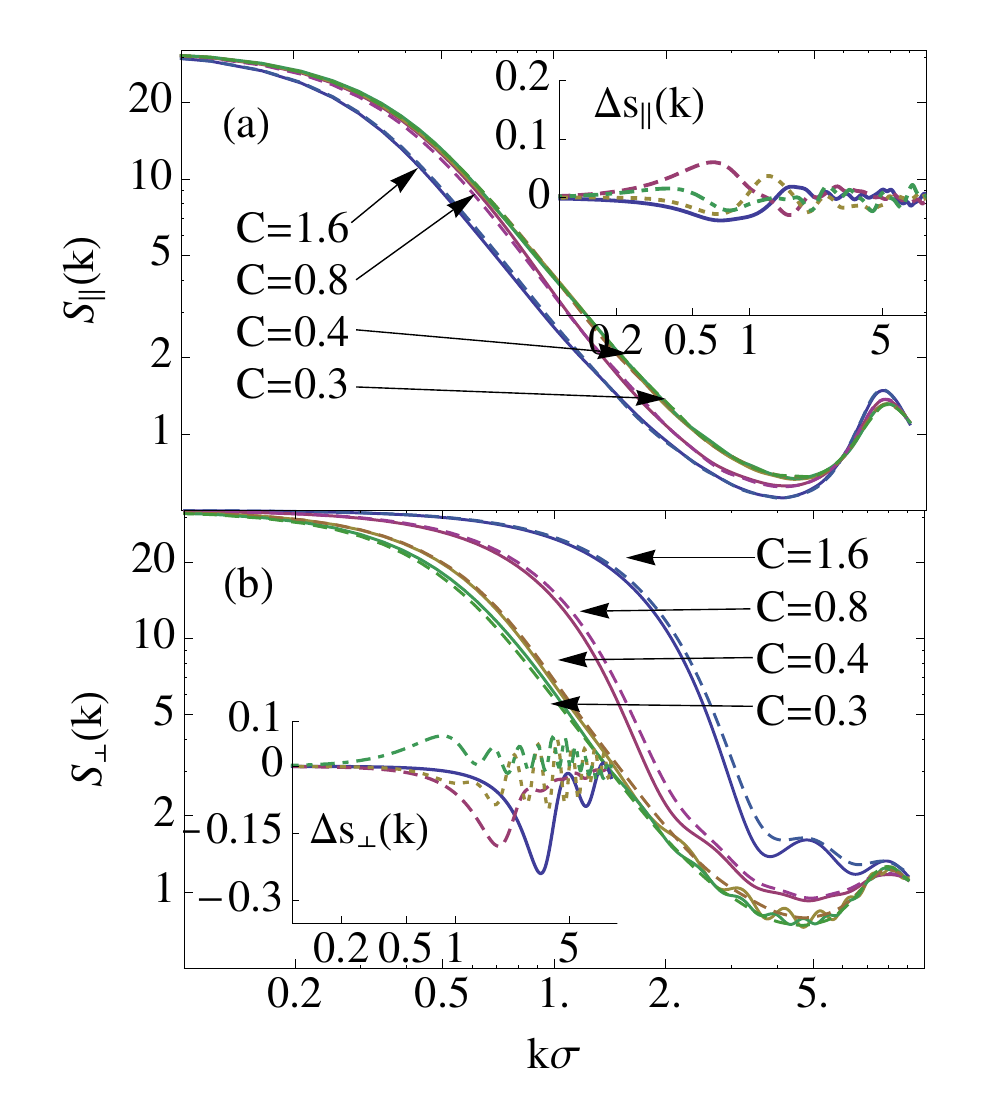}}
\caption{Both the (a) planar and (b) the perpendicular component of the static structure factor show differences when measured in q2D-LB (solid lines) and 3D-LB simulations (dashed lines). The polymer contains $32$ beads. The insets show the differences defined in Eq.~(\ref{eq:littlesk}) at different $C$ as a function of $k\sigma$. The lines in the insets correspond to $C=1.6$ (solid), $0.8$ (dashed), $0.4$ (dotted) and $0.3$ (dot-dashed).}
\label{fig:Sk3296LBLD}
\end{figure}

Y. von Hansen {\it et al.} found that as a polymer's mean distance to a wall is decreased, there is a slow transition from Zimm to Rouse dynamics in terms of hydrodynamic screening effects~\cite{HHN11}. Their observation leaves room for the possibility that even static quantities such as the static structure factor could be affected by the presence of confining walls. We investigate this in Fig.~\ref{fig:Sk3296LBLD}, which shows the planar ($S_\parallel(k)$, Eq.~(\ref{eq:Sktconf}a)) and perpendicular ($S_\perp(k)$ Eq.~(\ref{eq:Sktconf}b)) components at different levels of confinement for q2D-LB fluid confined between two no-slip walls (solid lines in Fig.~\ref{fig:Sk3296LBLD}(a) and (c)) and 3D-LB fluid (dashed lines in Fig.~\ref{fig:Sk3296LBLD}(a) and (c)). Two parallel plates confine the chain in both cases according to Eq.~(\ref{eq:Umw}). A comparison between the q2D-LB and the 3D-LB fluid (a box of size $(52\Delta x)^3$ with PBCs) allows us to ascertain whether an effect due to direct momentum absorption by the no-slip walls in the $z$ direction can be seen in the static structure factor. Figures~\ref{fig:Sk3296LBLD}(a) and (b) show the differences
\begin{equation}\label{eq:littlesk}
\Delta s(k) = \frac{S_\mathrm{q2D,LB}(k)-S_\mathrm{3D,LB}(k)}{S_\mathrm{3D,LB}(k)}
\end{equation}at different levels of confinement in the plane of the walls ($\parallel$) and perpendicular to it ($\perp$). Differences up to $4\%$ ($\parallel$) to $24\%$ ($\perp$) are visible at strong confinement ($C=1.6$, solid lines), and at the opposite end ($C=0.3$, dot-dashed line), the maximum differences in the planar and perpendicular data are $4\%$ and $7\%$. These observations support the finding that $R_{g,\perp}$ approaches the unconfined limit more slowly than $R_{g,\parallel}$ in Fig.~\ref{fig:Rg}. The no-slip wall has an observable effect on chain statistics in the perpendicular direction at $C=0.8$ and $C=1.6$ for for $k\sigma \geq 1.5$. The largest deviations in the parallel direction occur at $C=1.6$ for $0.25 \leq k\sigma \leq 1.0$, which corresponds to lengths $(1.7 R_g, 7 R_g)$. However, these are not statistically significant as the maximum statistical errors for that range of $k\sigma$ in the differences are $0.06$ at $C=1.6$ (solid line) and $0.04$ at $C=0.8$ (dashed line) in the inset of Fig.~\ref{fig:Sk3296LBLD}(a).
\begin{figure}
\centerline{\includegraphics[width=1.0\columnwidth]{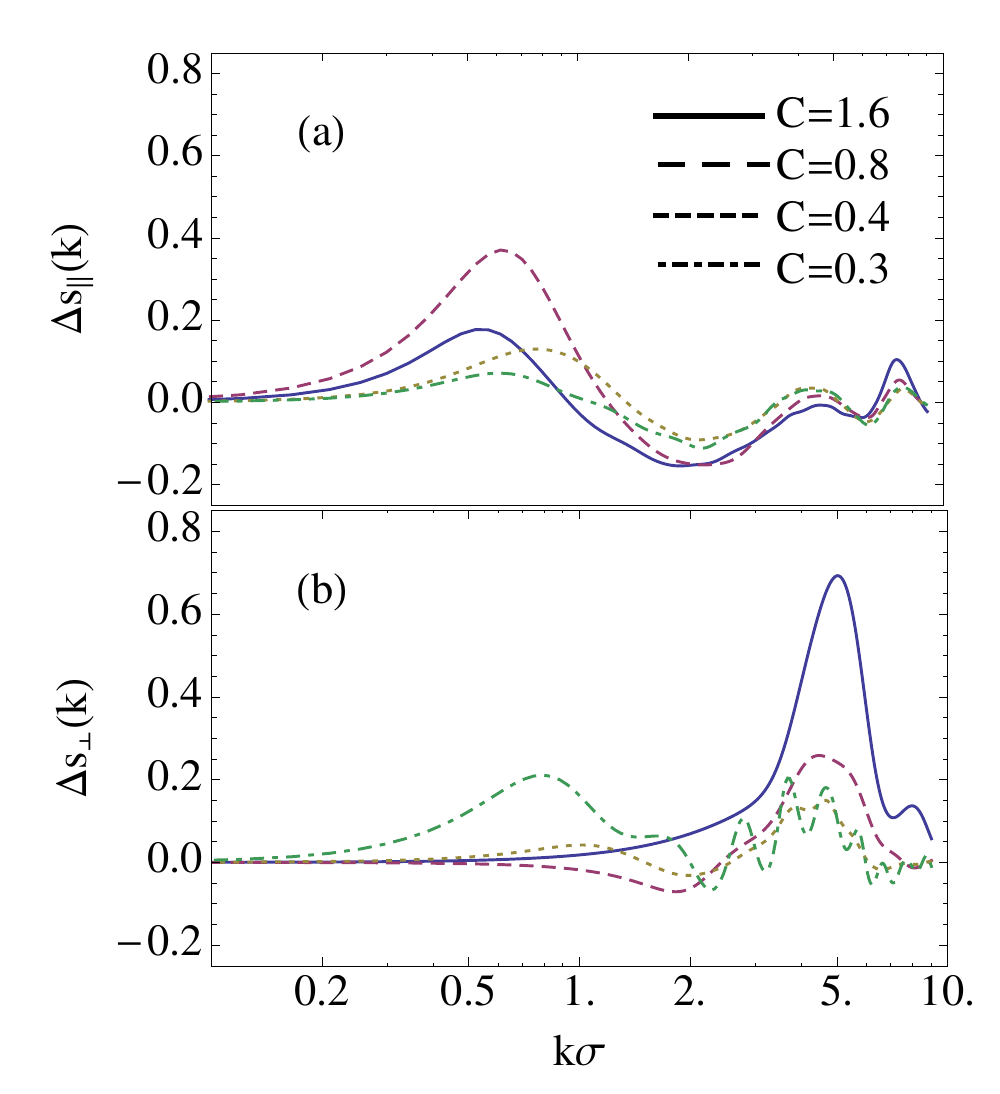}}
\caption{The normalized difference in the (a) planar and (b) perpendicular static structure factor between q2D-LB and LD simulations at different degrees of confinement (Eq.~(\ref{eq:C})). Structural differences can be seen at values of $k\sigma$ corresponding to the monomer size ($k\sigma \approx 6.3$), the maximum bond length $k\sigma=4.2$, the spacing between plates at $C=1.6$ (solid line at $k\sigma \approx 3.5$), the distance between next-nearest neighbor monomers ($k\sigma \approx 2$) and twice the radius of gyration ($k\sigma=0.7,\ldots,0.9$). } 
\label{fig:N32Skq2DvsLang}
\end{figure}

Next, we shall ascertain whether the static structure factor shows any differences due to the presence of hydrodynamic interactions. This is done by comparing results from q2D-LB and LD simulations, which we comment on first. The monomer structure in LB is stiffer due to the composite nature of monomers of Fig.~\ref{fig:schematic}.

Figure~\ref{fig:N32Skq2DvsLang} shows differences in the static structure factor, but between q2D-LB and LD simulations, {\it i.e.} $\Delta s(k) = (S_\mathrm{q2D,LB}(k)-S_\mathrm{LD}(k))/S_\mathrm{LD}(k)$. The planar data in panel (a) tell that the Langevin chain frequents configurations for which $1.3 < k\sigma < 3$ more than the LB chain (maximum statistical error in $\Delta s_\parallel(k)$ is $0.04$ for $C=1.6$ and $0.03$ for $C=0.8$) and the situation is reversed for $0.4 < k\sigma < 0.9$ (maximum error $0.09$ for $C=1.6$ and $0.05$ for $C=0.8$). The former range of $k\sigma$ corresponds to $R_g$ and the latter to $2 R_g$. The differences become smaller as $C$ tends closer to zero (dotted and dot-dashed lines) as is expected for chains in bulk. The perpendicular data at $C=1.6$ in panel (b) reveal the local stiffness of bonds in the LB fluid ($k\sigma= (2\pi/(2.25\,\mathrm{nm}))\sigma \approx 4.2$). Again, as $C$ tends to zero, the difference in $s_\perp(k)$ becomes smaller.

\begin{figure}
\centerline{\includegraphics[width=0.975\columnwidth]{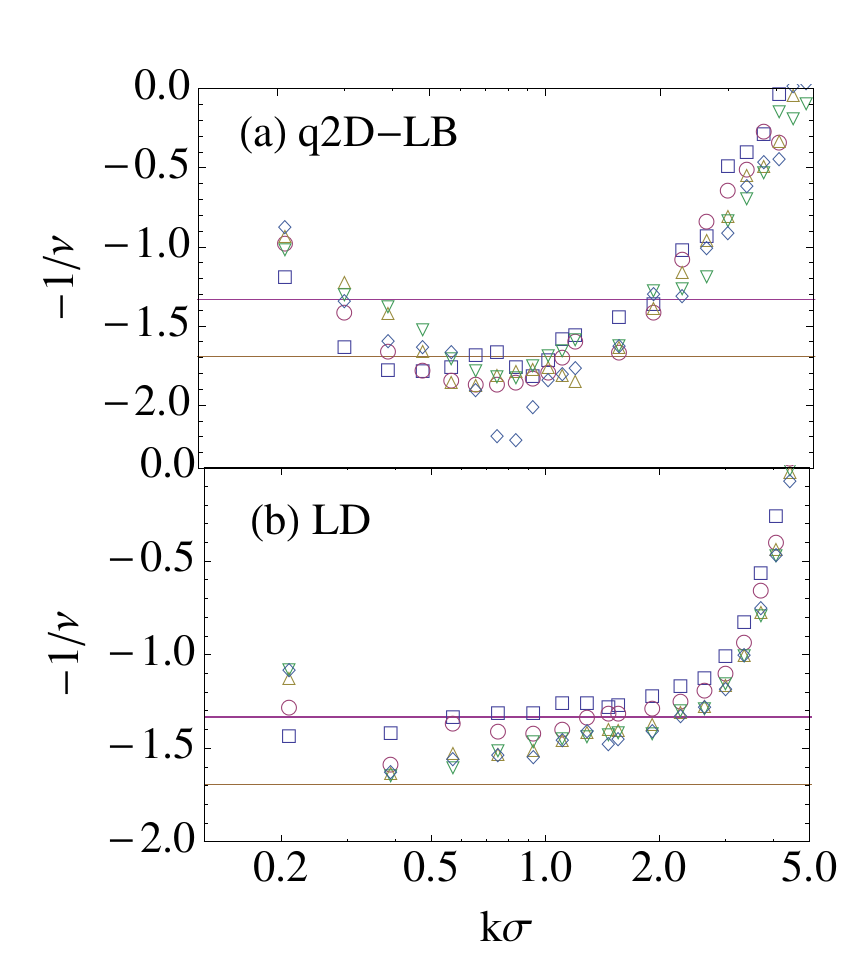}}
\caption{Exponent $-1/\nu$ based on the relation $S_\parallel(k) \sim k^{-1/\nu}$ for a chain at levels of confinement $C=2.0$ (squares), $0.9$ (circles), $0.5$ (up triangles), $0.3$ (down triangles) and $0.2$ (diamonds). Panels are for a $96$-bead chain from (a) LB and (b) LD simulations. The top horizontal line indicates the 2D scaling exponent $-1/\nu_\mathrm{2D}=-4/3$ and the bottom line the 3D exponent $-1/\nu_\mathrm{3D}= -1.7$.}
\label{fig:alphaofSk}
\end{figure}

With the possibility of scaling in the form of a power law, we plot $-1/\nu = \partial [\log S(k)]/\partial [\log k]$ as a function of $k$ in Fig.~\ref{fig:alphaofSk}. We entertain the questions if and which scaling exponent the planar static structure factors exhibit. In approaching the 2D limit, $C$ would tend to infinity and $1/\nu=4/3$ and in the 3D limit, $C\rightarrow 0$ and $1/\nu \approx 1.7$ for a long chain in the scaling region. Figure~\ref{fig:alphaofSk} has two panels showing $-1/\nu$ for $N=96$ extracted from the planar structure factor, $S_\parallel(k)$, based on q2D-LB and LD simulations. The LB data have a decay exponent larger than $1/\nu=1/0.6$ in the anticipated scaling region whereas Langevin simulations in Fig.~\ref{fig:alphaofSk}(b) follow the line $-1/0.75$ corresponding to 2D-like scaling for $0.8 < k\sigma < 2$ especially at the highest degree of confinement ($C=2.0$, hollow squares in (b)). That is, the static structure in the presence of HIs suggests 3D dynamics whereas strong confinement in LD produces effectively 2D dynamics for the same value of $C$. The q2D-LB data comes closer to the $-1/\nu_\textrm{2D}$ line for $C=3.1$ (not shown), which is the most confined case we can access. The monomer size corresponds to $k\sigma=2\pi$ and the radius of gyration to $k\sigma=0.8$. Moreover, the unconfined limit $C\rightarrow 0$ is approached differently by the two types of dynamics: LB data do not exhibit clear 2D scaling in the anticipated scaling region, but indicate $1/\nu$ to decrease from about $2$ to $1/0.6$ as $C \rightarrow 0$ whereas LD data transition from 2D ($1/\nu=1/0.75$) to 3D scaling ($1/0.6$) as $C\rightarrow 0$.

\subsection{Dynamic scaling}
Equilibrium polymer dynamics in bulk is described by the theory of dynamic scaling~\cite{doiedwards,deG79}. The scaling prediction of Eq.~(\ref{eq:Sktscaling}) in the presence of HIs can be tested by plotting $S(k,t)/S(k,0)$ versus $(k\sigma)^z (t/\tau_\mathrm{LJ})$ for different $k$, which should all collapse onto a single curve, {\it i.e.} $F$ of Eq.~(\ref{eq:Sktscaling}), with $z=3$ in three dimensions and $z=2$ in two dimensions~\cite{FPVA-N03}. The corresponding theoretical values in the absence of HIs are $z=3.7$ and $z=3.33$~\cite{doiedwards,deG79}.

\begin{figure*}
\centerline{\includegraphics[width=1.0\textwidth]{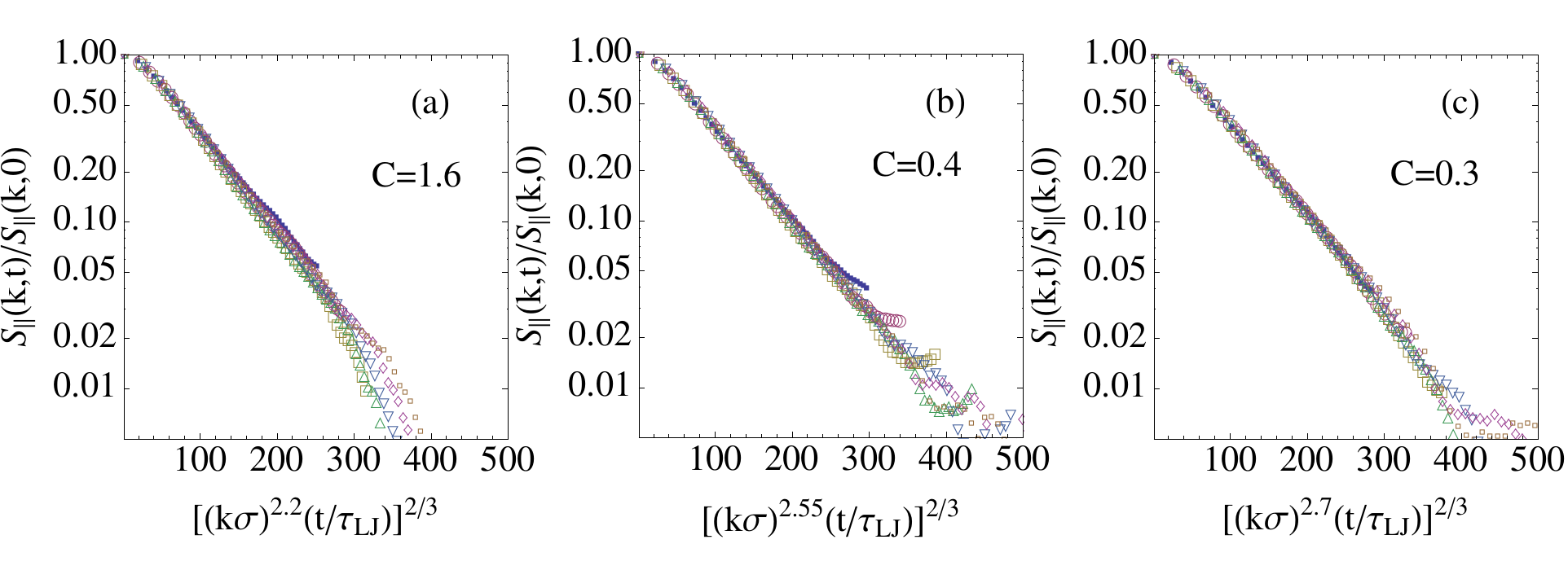}}
\caption{Dynamic scaling for $N=32$ in the presence of HIs along ${\bf k}=k({\bf e}_x+{\bf e}_y)/\sqrt{2}, k\sigma\,\in\,(1.8,2.7)$, {\it i.e.} parallel to the confining walls. As the level of confinement decreases from (a) $C=1.6$ to (b) $C=0.4$ and to (c) $C=0.3$, the scaling exponent increases from $z=2.2\pm 0.1$ to $z=2.7\pm 0.1$.}
\label{fig:N32Skt}
\end{figure*}

\begin{figure}
\centerline{\includegraphics[width=0.9\columnwidth]{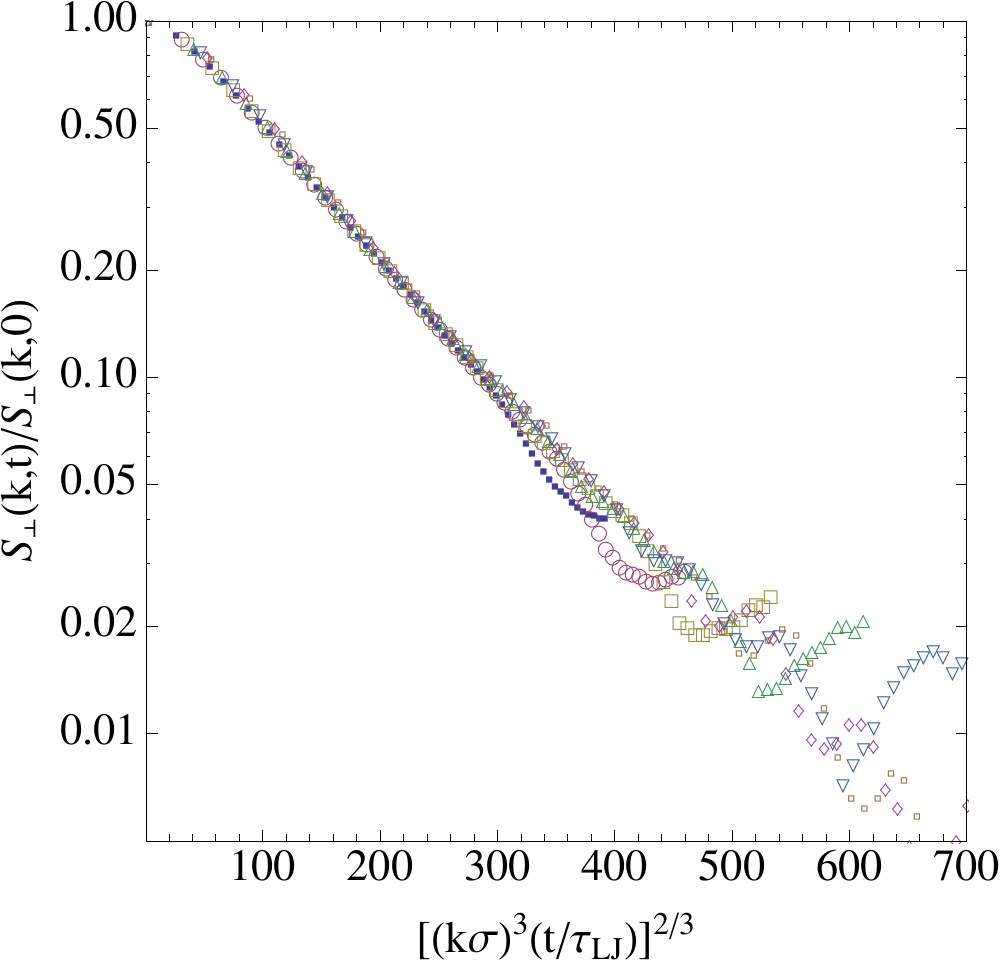}}
\caption{For plates spaced $C=0.3$ apart, the dynamic structure factor in LB fluid for $N=32$ exhibits 3D scaling behavior in the direction perpendicular to the plate.}
\label{fig:N32SktH3perp}
\end{figure}

Since momentum conservation is broken in the $z$ direction in our q2D system, one cannot expect either result to hold as such. It is therefore interesting to see what the transition between 3D and 2D scaling behavior looks like. As in the static case, we treat the $xy$ plane and the perpendicular $z$ direction separately. We find $S_\parallel(k, t)$ for $N=32$ with HIs to exhibit values of $z$ that are close to the known 2D ($z=2$) result in strong confinement and approach the 3D result ($z=3$) as the level of confinement decreases. Evidence of this transition is provided in Fig.~\ref{fig:N32Skt}, which shows scaling collapses for $S_\parallel(k,t)$ in which $z$ increases from $2.2$ at $C=1.6$ to $2.7$ at $C=0.3$. The $C\rightarrow 0$ limit was studied in Ref.~\onlinecite{ODKA11} and gives $z=3$. The collapses are plotted as functions of $(k^z t)^{2/3}$ as it has been shown theoretically that $\log F(x) \sim x^{2/3}$ for $k^3 k_\mathrm{B} T t/(6 \pi \eta) \gg 1$ in free space~\cite{WHR97}. Since the right-hand side of the inequality is at least $10$ for our parameters, we opted to use this form in the scaling plots. When probed perpendicular to the wall using ${\bf k} = k{\bf e}_z$, $S_\perp(k,t)/S_\perp(k,0)$ do not collapse for any $z$ at $C=1.6$ {\it i.e.} at plate separation $\approx R_g/2$ (not shown), but at $C=0.3$ (plate separation of $\approx 3 R_g$) plotted in Fig.~\ref{fig:N32SktH3perp}, we do observe a reasonable data collapse for $N=32$ with a scaling exponent of $z=3$ indicating 3D behavior. As Langevin simulations for long chains are feasible, we produced scaling collapses of $S_\parallel(k,t)$ for $N=96$ at $C=1.8$. We found the chain to scale very clearly with the 2D scaling exponent of $z=2+1/0.75=3.33$ for $k\sigma \in (1.4,2.8)$ (not shown).

\subsection{Diffusion}
Recent experiments by Strychalski {\it et al.}~\cite{SLC08} and by Tang {\it et al.}~\cite{TLTJCD10} have shown the degree of confinement to be reflected in the chain's planar diffusion coefficient as $D_\parallel \sim C^{-\alpha}$ ($0.4<C<25$) and ($0.3<C<21$) in aqueous solution. Moreover, their respective findings $\alpha \approx 0.48\pm 0.05$ (as averaged over Strychalski's results for different chain lengths) and $\alpha = 0.49\pm0.01$ are in agreement with previous experiments~\cite{HBD07,BMHD06}, but in contradiction with the predictions of blob theory that gives $\alpha = 2/3$~\cite{doiedwards,deG79}. However, Usta {\it et al.}~\cite{UBL05} found agreement with blob theory in slit-like confinement using their Lattice-Boltzmann scheme. Their LB scheme differs from ours in that we need not add external Langevin noise to our chain as we have consistently calibrated the hydrodynamic radius of our monomers and the coupling between the monomers and the LB fluid~\cite{ODKA11}. 

\begin{figure}
\centerline{\includegraphics[width=1.0\columnwidth]{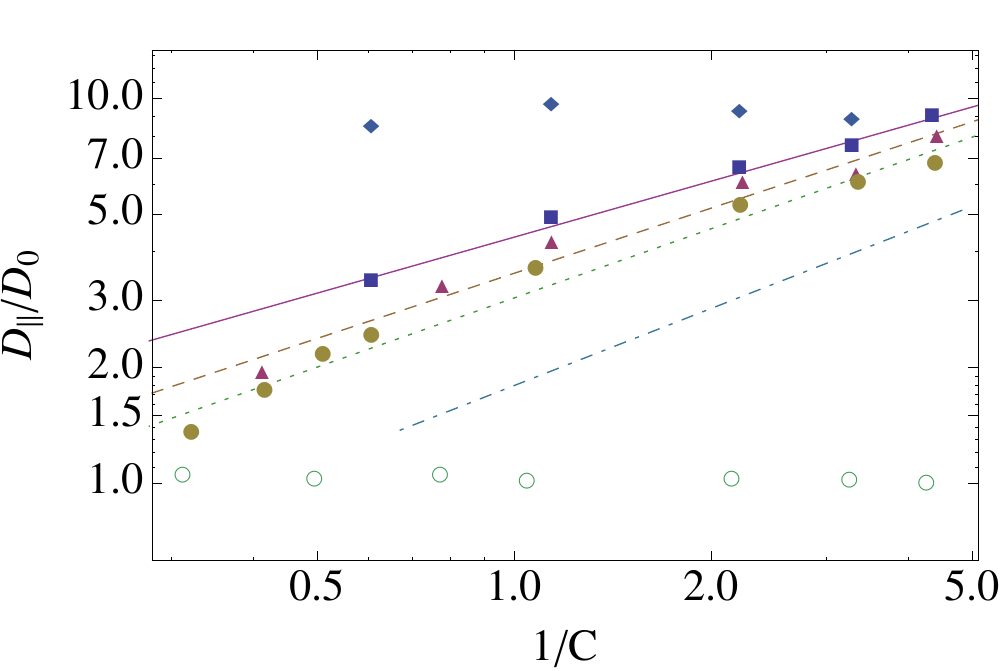}}
\caption{The scaled planar center-of-mass diffusion coefficient as a function of $1/C$ in q2D-LB for $N=32$ (solid squares), $N=64$ (solid triangles) and $N=96$ (solid circles), in 3D-LB for $N=32$ (solid diamonds) and LD for $N=96$ (hollow circles). The fits to q2D-LB data give $D_\parallel \sim C^{-0.48\pm 0.04}$ (solid line, $N=32$), $\sim C^{-0.56\pm 0.04}$ (dashed line, $N=64$) and $\sim C^{-0.60\pm 0.04}$ (dotted line, $N=96$). The errors are of the size of the symbols. The dot-dashed line with a slope of $2/3$ corresponds to the prediction based on blob theory. The scaling factor $D_0 = k_\mathrm{B} T (N\xi)^{-1}$ is the diffusion coefficient for $N=96$ based on Rouse dynamics.}
\label{fig:DN32}
\end{figure}

We define the in-plane center-of-mass diffusion coefficient of the chain through the mean squared displacement as
\begin{eqnarray}
D_\parallel = \displaystyle\lim_{\vert t-t_0\vert \rightarrow \infty} \frac{\langle ({\bf r}_{\parallel,\mathrm{cm}}(t) - {\bf r}_{\parallel,\mathrm{cm}}(t_0))^2\rangle}{4\,\vert t - t_0 \vert},\label{eq:Dmsd}
\end{eqnarray}where ${\bf r}_{\parallel,\mathrm{cm}}$ is the 2D vector $(x_\mathrm{cm},y_\mathrm{cm})$. Figure~\ref{fig:DN32} shows $D_\parallel$ in log-log scale as a function of $1/C$ for $N=32$, $64$ and $96$. In general, we observe a reduction in the diffusivity as the level of confinement is increased and for a given value of $C$, the longer a chain is the slower it diffuses.

We find a single power law exponent to fit our $\log D_\parallel$ vs. $\log\, (1/C)$ data nicely over a decade starting from $1/C=0.5$. However, by including data points in the region $1/C<0.5$, we get $D_\parallel \sim C^{-0.48 \pm 0.04}$ for $N=32$ (solid squares), $\sim C^{-0.56\pm 0.04}$ for $N=64$ (solid triangles) and  $\sim C^{-0.60\pm 0.04}$ for $N=96$ (solid circles). Linear interpolation suggests $C^{-0.66\pm 0.05}$ for $N\rightarrow \infty$, which agrees with de Gennes' theory. 

The parallel-plate arrangement is suitable for testing the effect of direct momentum absorption by the no-slip walls in the $z$ direction on a dynamic quantity such as the chain diffusivity. Interestingly, once the no-slip walls are removed (MD walls are still kept in place) and the LB fluid is then contained in a periodic box of size $(52\Delta x)^3$, $D_\parallel$ for $N=32$ (solid diamonds) becomes independent of $C$. Even at $1/C\approx 3.6$, the effect of q2D HIs and the non-conserved $z$ momentum on $D_\parallel$ is still $14\%$ ($D_{\parallel,\mathrm{q2D}}/D_{\parallel,\mathrm{3D}}\approx 0.86$).
The LD simulations for $N=32$ (not shown for clarity) and $96$ (hollow circles) are also unaffected by a change in $C$, which together with the LB simulations reveals the importance of the no-slip boundary condition with respect to the experimental observations. We note that even at the strongest confinement ($1/C \approx 0.3$), $D_\parallel(\mathrm{LB},N=96)/D_\parallel(\mathrm{LD},N=96) \approx 1.3$ and thus, we concur with the conclusions by von Hansen~{\it et al.}~\cite{HHN11} regarding the broadness of the transition from Zimm to Rouse dynamics in the vicinity of a wall.

%%%%%%%%%%%%%%CONCLUSIONS %%%%%%%%%%%%%%%%%%
\section{Conclusions}
We have simulated solvated coarse-grained polymers between two parallel plates of varying separation with (The Lattice-Boltzmann Method) and without (Langevin Dynamics) hydrodynamic interactions. As our numerical method is not applicable for studying the Odijk regime, we have only used plate separations that correspond to a regime where the de Gennes blob theory should hold. The parallel-plate system is unique due to the form of pairwise hydrodynamic interactions~\cite{Diamant09} and as the polymer moves almost in two dimensions when the plate separation is comparable to or smaller than the polymer's free-space radius of gyration. For fixed MD confinement, we found hydrodynamic interactions to change the planar monomer distribution about the chain CM slightly as the dimensionality of the fluid was changed from q2D to full 3D. For levels of confinement $C=0.3-2.0$, the Flory exponent based on the static structure factor attained values that varied between the free-space 2D and 3D values depending on the length scale in question. The planar diffusion coefficient for the CM of the chain, $D_\parallel$, decreased as a function of increasing confinement in agreement with experiments~\cite{SLC08,TLTJCD10,HBD07,BMHD06}. For the limited range of $C$ accessible in our simulations, this suggests the scaling behavior to be due to the presence of no-slip walls alone and electrostatic interactions are not needed to account for the observed reduction. However, the value of the scaling exponent seemed to agree with experiments~\cite{SLC08,TLTJCD10,HBD07,BMHD06} for short chains and asymptotically with de Gennes' prediction. As experiments are carried out close to the long-chain limit and are unaffected by finite-size effects, more studies are needed to ascertain the role of electrostatics in the scaling of the planar diffusion coefficient. Langevin dynamics (LD) simulations do not capture the reduction in $D_\parallel$ as a function $C$ for which reason they are not a suitable tool for the study of dynamic phenomena in a confined environment. However, the LD results are qualitatively similar to confinement imposed only on the polymer, but not the solvent. We have also characterized the differences in the static structure factor between different types of confinement and dynamics. Finally, we observed a continuous transition in the polymer's dynamic scaling exponent from its value in 2D to the one in 3D as a function of decreasing level of confinement in the de Gennes and weak-confinement regime.

\begin{acknowledgments}
This work was supported by the Academy of Finland through its COMP Centres of Excellence Program (project no. 251748), by the Natural Science and Engineering Council of Canada and by SharcNet. We also wish to thank CSC, the Finnish IT center for science, for allocation of computer resources. S.T.T.O. thanks the Emil Aaltonen Foundation for funding.  This research was supported in part by the National Science Foundation under Grant No. NSF PHY11-25915.
\end{acknowledgments}

%%%%%%%%%%%BIBLIOGRAPHY%%%%%%%%%%%%%%%%%%%%

\end{document}